\documentclass[aps,prb,reprint,superscriptaddress,showpacs]{revtex4-1}
\usepackage{color}
\usepackage{graphicx}
\usepackage[american]{babel}
\usepackage[utf8]{inputenc}
\usepackage[T1]{fontenc}
\usepackage{bbm,ae,aecompl}
\usepackage{amsmath}
\usepackage{amssymb}
\usepackage{amstext}
\usepackage{amsthm}
\usepackage{latexsym}
\usepackage{verbatim}
\usepackage{natbib}
\usepackage{array}

\usepackage{ulem}   

\begin{document}

\title{Transfer of spectral weight across the gap of Sr$_2$IrO$_4$ induced by La doping}
\date{\today}

\author{Véronique Brouet}

\author{Joseph Mansart}
\affiliation{Laboratoire de Physique des Solides, Universit\'{e} Paris-Sud, UMR 8502, B\^at. 510, 91405 Orsay, France}

\author{Luca Perfetti}

\author{Christian Piovera}
\affiliation{Laboratoire des Solides Irradi\'{e}s, Ecole polytechnique, 91128 Palaiseau cedex, France}

\author{Ivana Vobornik}
\affiliation{CNR-IOM, TASC Laboratory, AREA Science Park - Basovizza, I-34149 Trieste, Italy}


\author{Patrick Le F\`evre}

\author{François Bertran}
\affiliation{Synchrotron SOLEIL, L'Orme des Merisiers, Saint-Aubin-BP 48, 91192 Gif sur Yvette, France}

\author{Scott C. Riggs}
\author{M. C. Shapiro}
\author{Paula Giraldo-Gallo}
\author{Ian R. Fisher}
\affiliation{Geballe Laboratory for Advanced Materials and Department of Applied Physics, Stanford University, Stanford, California 94305-4045, USA}

\begin{abstract}
We study with Angle Resolved PhotoElectron Spectroscopy (ARPES) the evolution of the electronic structure of Sr$_2$IrO$_4$, when holes or electrons are introduced, through Rh or La substitutions. At low dopings, the added carriers occupy the first available states, at bottom or top of the gap, revealing an asymmetric gap of 0.7eV, in good agreement with STM measurements. At further doping, we observe a reduction of the gap and a transfer of spectral weight across the gap, although the quasiparticle weight remains very small. We discuss the origin of the in-gap spectral weight as a local distribution of gap values. 
\end{abstract}
\maketitle

The reaction of an insulator to doping can reveal many things on its underlying structure. For a band insulator, a simple shift of the chemical potential into bands that were previously completely filled or empty can be expected. If more complex electronic correlations are involved, the formation of in-gap states and/or large transfer of spectral weight across the gap could occur \cite{GeorgesRMP96,MeindersPRB93}. This problem received a lot of attention in the case of cuprates \cite{DamascelliRMP03,KMShenPRL04}. It is very interesting to investigate this in Sr$_2$IrO$_4$, whose insulating nature is not completely understood. The idea driving the field for a few years is that strong spin-orbit coupling reshapes the electronic structure in a way that enhances correlations \cite{BJKimPRL08}. Especially, a narrow half-filled band of J=1/2 character forms at the Fermi level, which could be split by correlations to form a Mott insulator. The exact role of the antiferromagnetic (AF) ordering observed below 240~K \cite{YeCaoPRB13} in the insulating nature is still debated \cite{MartinsPRL11,AritaPRL12}. On the other hand, applying pressure, even up to 55~GPa, was not able to close the gap \cite{ZoccoJCondMat14}, which is surprising for a Mott insulator. In this context, it would be interesting to know more about the electronic structure of doped phases, which exhibit metallic-like behaviors even for rather low doping rates \cite{GeCaoPRB11,QiCaoPRB12,LeeTokuraPRB12,ClancyPRB14}. 
\begin{figure}[b]
\centering
\includegraphics[width=0.47\textwidth]{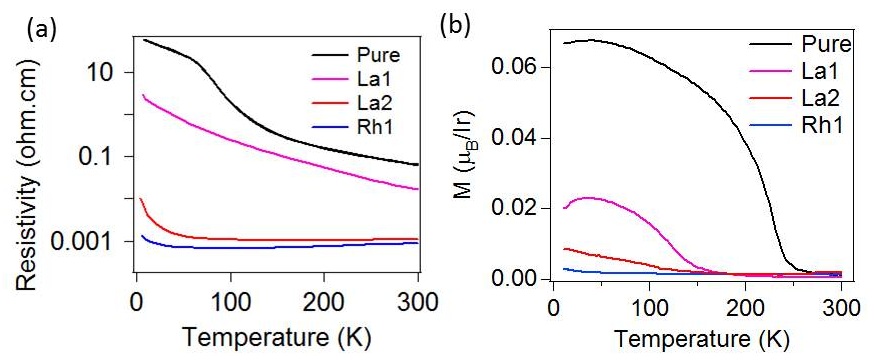}
\caption{(a) In-plane resistivity in zero magnetic field and (b) magnetization at 1T, measured for the main samples studied in this paper : pure Sr$_2$IrO$_4$, La1 (1.5\% La), La2 (4\% La) and Rh1 (15\% Rh).
\label{Fig_Samples}}
\end{figure}
\vspace{0.3cm}
\begin{figure}[htb]
\centering
\includegraphics[width=0.45\textwidth]{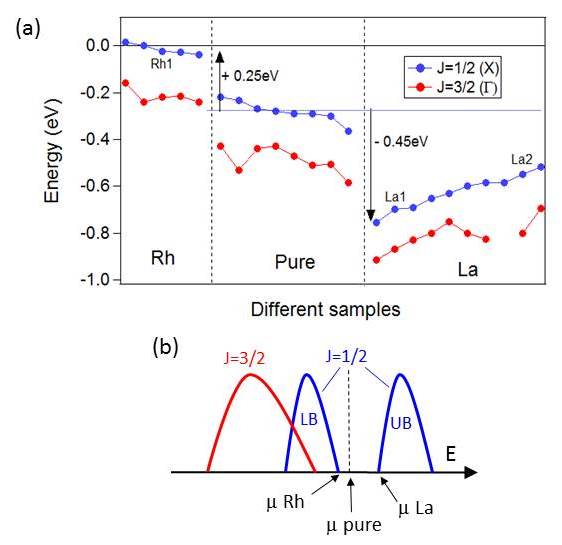}
\caption{(a) Energy positions of the J=1/2 peak at X (blue points) and J=3/2 peak at $\Gamma$ (red points) measured around 50K for different samples of Sr$_2$IrO$_4$ (middle), Sr$_2$IrO$_4$ doped with Rh (left) and La (right). The axis roughly corresponds to different dopings, ranging from 10 to 15$\%$ in the case of Rh and 1 to 4$\%$ in the case of La. (b) Sketch of the shift of the chemical potential with initial doping. As the J=1/2 band is gapped, it is divided into a lower band LB and an upper band UB.} 
\label{Fig_mushift}
\end{figure}

Recently, a few Angle Resolved Photoemission (ARPES) studies of doped iridate phases were reported, but they lead to quite a confusing picture. For Rh substitutions, which results in effective hole doping \cite{ClancyPRB14}, a metallic-like state was observed, albeit with residual pseudogaps instead of well defined quasiparticles (QP) peaks \cite{CaoDessauCondMat14}. Another kind of metallic state was observed by evaporating K on the surface of Sr$_2$IrO$_4$, which presumably dopes electrons into it \cite{KimScience14}. In this case, well defined QP were observed, with strong momentum and temperature dependences that resemble those found in the cuprates and even possible signs of superconductivity\cite{KimCondMat15,YanCondMat15}. Very recently, results similar to intermediate K coverage were reproduced in La doped Sr$_2$IrO$_4$ \cite{deLaTorre124}. In these cases, a large Fermi Surface (FS) containing $\sim$ 1$\pm$$x$ electrons was apparently observed ($x$ is the number of added carriers), although the spectral weight along it could be strongly modulated. In contrast, La doping in Sr$_3$Ir$_2$O$_7$ apparently produces a small FS containing only the added \textit{x} electrons\cite{TorrePRL14} with a progressive reduction of the gap \cite{HeNatMat15}. Clearly, the phenomenology of the metal-insulator transition in this system is still an intriguing and open problem.    

To better understand this, we present here a study of both hole and electron doped Sr$_2$IrO$_4$, using respectively Rh and La substitutions. Upon doping, we observe a nearly rigid shift of the band structure, towards the Fermi level for Rh and to higher binding energies for La. This reveals an electronic gap of 0.7eV, larger than what is usually assumed, because it is significantly larger on the unoccupied side. The spectral weight at the Fermi level follows a distribution in k-space that is surprisingly different for Rh (circular-like) and La (squarish-like). By examining the nature of the in-gap weight, we attribute this to a local distribution of gap values. This is in good agreement with observation by STM that the gap can be strongly reduced near defects \cite{DaiPRB14,OkadaNatMat13}. At the largest La doping that we could synthesize, we observe a small QP peak emerging at the M point.

The samples were prepared using a self-flux method, as reported in \cite{KimScience09}. Fig. \ref{Fig_Samples} shows examples of the resistivity and magnetization measured for the main samples that will be used in this study. Their exact doping was estimated by Energy Dispersion X-ray analysis to be 1.5$\%$ for La1, 4$\%$ for La2 and 15$\%$ for Rh1. Upon La and Rh dopings, the resistivity drops, as observed by other groups \cite{GeCaoPRB11,QiCaoPRB12,LeeTokuraPRB12}, with a slight upturn at low temperatures. The magnetic transition is suppressed by doping and it is not detectable anymore in the La2 and Rh1 samples. ARPES experiments were carried out at the CASSIOPEE beamline of SOLEIL synchrotron and the APE beamline of ELETTRA synchrotron, with a SCIENTA R-4000 analyser and an overall resolution better than 15meV.

\begin{figure*}[t]
\centering
\includegraphics[width=1\textwidth]{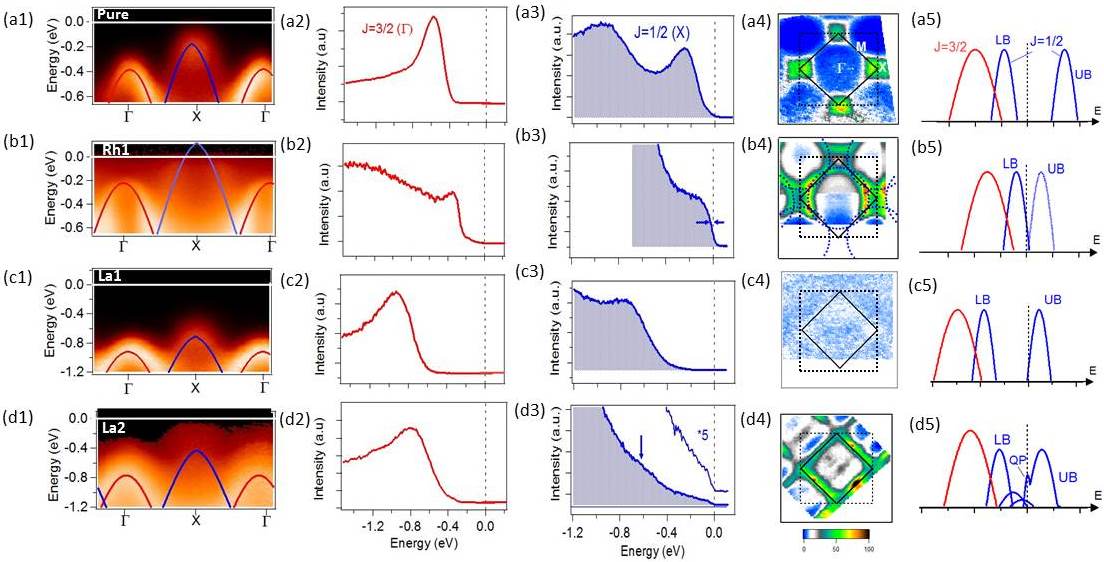}
\caption{First column : Energy-momentum plots in the $\Gamma$X direction. Red and blue lines are guides for the eye for the dispersion of J=3/2 and J=1/2 bands. Second column : Energy Distribution Curve (EDC) at $\Gamma$ for the J=3/2 band. Third column : EDC at X [k$_F$ for (b3)] for the J=1/2 band. Fourth column : Spectral weight integrated at -20meV in a 50meV window. The black square corresponds to the 2~Ir BZ and the dotted black square to the 1~Ir BZ. Fifth column : Sketch of the position of the main bands. The rows correspond to different samples (see Fig. 1) : (a) pure, (b) Rh1, (c) La1 and (d) La2. All data were acquired at 100eV and horizontal polarization, except for Rh1 acquired at 70eV and vertical polarization. The temperature was 50K. }
\label{Fig_GX}
\end{figure*}

Fig. \ref{Fig_mushift} summarizes the changes in energy for the top of the J=1/2 and J=3/2 bands in a series of compounds with different concentration of Rh (from 10$\%$ to 15$\%$), La (from 1$\%$ to 4$\%$) and several pure compounds. We will give details in Fig. \ref{Fig_GX} on how these positions are determined. They are sorted with respect to the J=1/2 value, which loosely corresponds to the doping level (we suspect that impurities and/or defects also play a role to fix the chemical potential position). Fig. \ref{Fig_mushift} evidences that, as soon as Rh and La are used, the bands suddenly shift in opposite directions to the bottom or top of the gap, as sketched in Fig. \ref{Fig_mushift}b. These shifts are not symmetric, being 0.2eV for Rh and -0.45eV for La. This is in very good agreement with the gap measured by STM \cite{DaiPRB14}, which is also asymmetric, of 0.15eV on the hole side and 0.47eV on the electron side. This confirms that the gap is larger than what is generally believed. For larger La doping, the bands shift back to the Fermi level, which we attribute to a reduction of the gap by up to 0.15eV. A similar behavior was observed recently in Sr$_3$Ir$_2$O$_7$ at low La dopings \cite{HeNatMat15}. For Rh, the gap lies on the unoccupied side, so that we cannot easily comment with ARPES on its size. 


In Fig. \ref{Fig_GX}, we detail this evolution. In the first column, we show the dispersion along the direction corresponding to the Ir-O-Ir bond, named $\Gamma$X in the unit cell containing 2~Ir \cite{CrawfordPRB94}. In Sr$_2$IrO$_4$, the t$_{2g}$ manifold is filled with 5 electrons and split into a quartet of J$_{eff}$=3/2 character and a doublet of J$_{eff}$=1/2 character \cite{BJKimPRL08}. In Density Functional Theory (DFT) calculations, both bands should cross the Fermi level. In reality, the J=3/2 band is pushed below E$_F$ and the J=1/2 band is gapped, as confirmed by previous ARPES studies in Sr$_2$IrO$_4$ \cite{BJKimPRL08,WangDessauPRB13} and Ba$_2$IrO$_4$ \cite{MoserNJP13,UchidaPRB14}. Along $\Gamma$X, a band of J=3/2 character is observed at $\Gamma$ and a band of J=1/2 character at X \cite{BJKimPRL08,WangDessauPRB13,MoserNJP13,UchidaPRB14}. Their dispersions are highlighted by red and blue lines for clarity (these guides are modeled on the DFT dispersions and then shifted to match the data). The lineshapes at $\Gamma$ and X are detailed by the Energy Distribution Curves (EDC) in the second and third columns. These peak positions are the ones reported in Fig. \ref{Fig_mushift}. 

In pure Sr$_2$IrO$_4$, the band approaching the closest from the Fermi level is the J=1/2 band at X, so that integration of the spectral weight at -20meV in a 50meV window gives patches of high intensity near X [Fig. 2(a4)]. Note that just the tail of the peak reaches E$_F$, so that this map is not a Fermi Surface. 

For a sample doped with 15$\%$ Rh (Rh1, second row), the bands are shifted almost rigidly towards the Fermi level, by approximately 0.2eV. While the J=3/2 band is still well below E$_F$ and completely filled, the J=1/2 band now crosses the Fermi level. There is however no detectable quasiparticle at k$_F$, as can be seen from the EDC in Fig. \ref{Fig_GX}(b3). In fact, the crossing is not even as sharp as expected for a Fermi Dirac function. From the point at half maximum on the leading edge (see arrows), one could define a residual pseudogap of $\sim$ 30meV. A similar behavior was reported in \cite{CaoDessauCondMat14}. We sketch this situation in Fig. \ref{Fig_GX}(b5), although we do not know if there is still a gap with UB. The spectral weight distribution map follows circles, sketched by blue dotted lines. If we neglect the pseudogap, this is the shape expected in DFT calculations for the Fermi Surface of the J=1/2 band of Sr$_2$IrO$_4$ near half doping. The circle radius k$_F$=0.63$\pm$0.04\AA$^{-1}$ defines an area corresponding to 0.95$\pm$0.1 electrons/Ir. This is compatible, within error bars, with a transfer of one hole per Rh to Ir. 

In Fig. \ref{Fig_GX}c and \ref{Fig_GX}d, we study 2 La dopings. At small doping (La1, in Fig. \ref{Fig_GX}c), the bands are strongly shifted down by about 0.45eV, in the opposite direction compared to Rh. As can be seen from the EDCs, the weight at the Fermi level is nearly zero and no clear features can be observed after integration. In this situation, there are enough added electrons to shift the Fermi level at the bottom of UB, but they do not modify the electronic structure of the insulator [Fig. \ref{Fig_GX}(c5)]. With further La doping, the gap reduces, as discussed in Fig. \ref{Fig_mushift}, which shifts the bands upward by ~150meV in La 2 (4th row). While this shift is almost rigid for the J=3/2 band, whose EDC shape is similar to that of La1, it is not for J=1/2. The J=1/2 peak is almost undetectable in the EDC of Fig. \ref{Fig_GX}(d3) (see arrow) and the lost weight appears to be transferred into the gap, as evidenced by the large tail of non zero intensity that extends up to the Fermi level. Although the weight at E$_F$ is very small, the EDC exhibits a small step there. The distribution of weight at the Fermi level in Fig. \ref{Fig_GX}(d4) is surprisingly well defined and forms a nearly perfect square along the 2~Ir BZ boundary. This is very different from the Rh case, although one could have expected similar shape from electron-hole symmetry. 

\begin{figure}[t]
\centering
\includegraphics[width=0.45\textwidth]{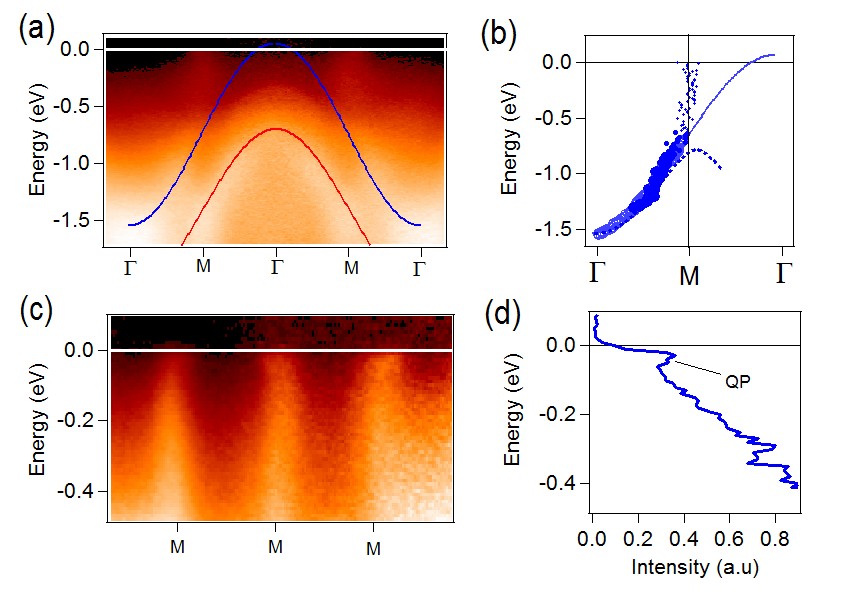} 
\caption{(a) Energy-momentum plot along $\Gamma$M measured for La2 at 50K. Blue (red) lines indicate the dispersion for J=1/2 (3/2) bands. (b) Dispersion extracted from (a) by MDC analysis. The size of the markers correspond to the amplitude of the peak. (c) Zoom near the Fermi level in the MM direction. (d) EDC at M extracted from (c). }
\label{Fig_La}
\end{figure}

The map nervertheless display higher intensity points at M (yellow color), which correspond to the appearance of a QP peak at M, as illustrated in Fig. \ref{Fig_La}. Along $\Gamma$M [Fig. \ref{Fig_La}(a)], we observe a J=3/2 hole band centered at $\Gamma$ (red line) and a deep electron band formed by the J=1/2 band (blue line, given by the DFT dispersion shifted down by 0.48 eV). The bottom of the band near $\sim$~-1.5eV is quite clear, but this band suddenly loses weight when reaching M. In Fig. \ref{Fig_La}b, its dispersion is extracted by fitting the Momentum Distribution Curves (MDCs) to Lorentzians. The peak positions are shown by blue markers and their amplitudes by the markers' size. The J=1/2 dispersion follows the blue line, up to binding energy of -0.6 eV, where the correlation gap opens, as sketched by the dotted line.  In the gapped region, we nevertheless observe reduced but non negligible intensity, as emphasized in Fig. 4c along MM direction. If fitted by MDC analysis, this in-gap weight gives rise to a nearly vertical line at M, from $\sim$-0.6eV up to E$_F$. Such a \lq\lq{}vertical dispersion\rq\rq{} often occurs when the MDC analysis picks up the tail of a broad peak \cite{VallaScience99}. It is particularly clear here due to the anomalously large weight in the tail of the J=1/2 EDC. Near the Fermi level, a small peak can be seen at M [Fig. \ref{Fig_La}(d)], which contrasts with the step observed at X in Fig. \ref{Fig_GX}(d3). As shown in ref.\cite{deLaTorre124}, this peak forms a small pocket around M with further La doping.

On the other hand, most of the weight picked up by the map in Fig. \ref{Fig_GX}d4 is dominated by the in-gap states. The fact that it forms a square very clearly indicates that the gap is tied to the 2~Ir BZ boundaries. Any interaction resulting in a 2~Ir unit cell could produce such a gap. In Sr$_2$IrO$_4$, the unit cell always contains 2 inequivalent Ir, because of the rotation of the oxygen octahedra \cite{CrawfordPRB94}. However, such a rotation cannot lift the degeneracy between the two bands at X. Moreover, in Ba$_2$IrO$_4$, where there is no such rotation, a similar gap is observed \cite{MoserNJP13,UchidaPRB14}. Interestingly, a distortion consisting of alternated elongation and squeezing along c of the oxygen octahedra has recently been proposed for Sr$_2$IrO$_4$ \cite{TorchinskyPRL15}. It could in principle produce such a behavior, although it is likely too small to produce such a large gap. The most likely candidate is of course the AF order that also produces a similar 2~Ir unit cell \cite{YeCaoPRB13}. Even though the gap does not close at T$_N$ in Sr$_2$IrO$_4$ \cite{MoonPRB09}, and though T$_N$ is strongly suppressed here by La doping, it is possible that AF fluctuations persisting above the transition are strong enough to maintain a pseudogap. In Ba$_2$IrO$_4$, a reduction of the gap above T$_N$ was suggested following similar reasonings \cite{UchidaPRB14}.

The most natural way to explain the large tail of the peak would then be \textit{a local distribution of gap values}. In Fig. \ref{Fig_GX}(d5), we sketch the 2 types of spectral weight transfer from the main LB occurring with La doping, (i) to the tail of the peak formed with LB having lower gap values and (ii) to the QP band. In fact, a rather similar picture emerges from STM studies. In ref. \cite{DaiPRB14}, the authors observed in Sr$_2$IrO$_4$ a gap as large as 0.62eV away from defects, but as small as 0.25eV near a defect, for example an oxygen vacancy. A similar behavior was observed in Sr$_3$Ir$_2$O$_7$ \cite{OkadaNatMat13}. This creates on average a tail into the gap, quite similar to the one we describe here. Indeed, even in pure Sr$_2$IrO$_4$, we do see a small weight inside the gap behaving very similarly to that found in the La doped compounds. It is very likely that this is why the chemical potential is pinned at 0.2eV above the gap, which is the position expected for the J=1/2 band in the DFT calculation. Korneta \textit{et al.} showed that at sufficient oxygen vacancies concentration (around 4$\%$), a metallic state forms \cite{KornetaPRB10}, suggesting that these ``defect states'' can be the precursor of a true metallic state.  

In the case of surface doped Sr$_2$IrO$_4$ (which yields electron doping, like La), a good metallic state, with well defined QP and circular FS, seems to be realized \cite{KimScience14}. Interestingly, the position of the J=3/2 band in this study is back to its original position near -0.5eV (see Fig. S1 of ref.\cite{KimScience14}). From the trend presented in Fig. \ref{Fig_mushift}, this case would then find its place as an extrapolation to larger La dopings, when the gap completely closes. The circular FS is analogous to the case of Rh we report here, restoring the idea of electron-hole symmetry. A difference is that there is no well defined QP in the case of Rh. They might develop at larger Rh dopings, but they could also be washed out by the disorder associated with the in-plane Ir/Rh substitutions. 

Finally, our picture for La doping is the following. At low doping levels, the La impurities generate a small amount of free negative charges that shift the Fermi Level to the bottom of the unoccupied band. By further increasing the La concentration, the gap is slightly reduced, at most by $\sim$150 meV for the dopings considered here. Locally, the La sites may further weaken the gap, creating a distribution of gap values responsible for large in-gap weight. Eventually, some percolation effect probably leads to the formation of conducting channels. The nature of correlation in this new state is one of the most interesting questions raised by the study of these iridates. The dispersive feature observed in all cases (pure, Rh and La doped) is well described by the unrenormalized DFT dispersion after appropriate shifts and should be viewed as a sizable incoherent feature. In the Rh case, it is difficult to separate this incoherent weight from the possibly emerging QP band at its top. In the La case, it becomes easier as the QP band forms on the other side of the gap. The transfer of spectral weight we observe as a function of doping is characteristic of a correlated system. The large size of the gap revealed by our study confirms STM results and is an important point that should be reproduced by any theory of this system. We also show that the gap can be highly sensitive to the local environment, which influences the way the doping proceeds. A better understanding of this effect should shed light on the nature of the gap and possible differences between chemical and surface doping, beyond the doping values both methods can presently reach.


\bibliography{bib_iridates}

\end{document}